\documentclass{elsart}
\usepackage{graphicx}
\usepackage{amsmath, amssymb}
\allowdisplaybreaks[4]
\newcommand{\vect}[1] {\mathbf{#1}}
\newcommand{\dif} {\mathrm{d}}
\newcommand{\up} {\uparrow}
\newcommand{\down} {\downarrow}
\newcommand{\I}{\mathrm{i}}
\newcommand{\E}{\mathrm{e}}

\newcommand{\brac}[1] {\langle #1\rvert}
\newcommand{\ket}[1] {\lvert #1\rangle}
\newcommand{\bracket}[2] {\langle #1 \rvert #2\rangle}

\begin{document}

\begin{frontmatter}
\title{Energetics of a strongly correlated Fermi gas}
\author{Shina Tan}
\ead{shina\_tan@yahoo.com}
\address{Institute for Nuclear Theory, University of Washington, Seattle, WA 98195-1550, U.S.A.}

\begin{abstract}
The energy of the two-component Fermi gas with the s-wave contact interaction
is a simple linear functional of its momentum distribution: 
$$E_\text{internal}=\hbar^2\Omega C/4\pi am+\sum_{\vect k\sigma}(\hbar^2 k^2/2m)(n_{\vect k\sigma}-C/k^4)$$
where the external potential energy is not included,
$a$ is the scattering length,
$\Omega$ is the volume, $n_{\vect k\sigma}$ is the average number of fermions
with wave vector $\vect k$ and spin $\sigma$,
and $C\equiv\lim_{\vect k\rightarrow\infty} k^4 n_{\vect k\up}
=\lim_{\vect k\rightarrow\infty} k^4 n_{\vect k\down}$. This result is a
\textit{universal identity}.
Its proof is facilitated by a novel mathematical idea, which
might be of utility in dealing with ultraviolet divergences in quantum field theories.
Other properties of this Fermi system, including the short-range structure of the one-body reduced density matrix and the pair correlation function, and the dimer-fermion scattering length, are also studied.
\end{abstract}

\end{frontmatter}

\section{\label{sec:introduction}Introduction}
It is a textbook fact that the energy of a noninteracting two-component Fermi gas is a
simple functional of its momentum distribution:
\begin{equation*}
E_\text{noninteracting}=\sum_{\vect k\sigma}\epsilon_{\vect k}n_{\vect k\sigma},
\end{equation*}
where $n_{\vect k\sigma}$ is the average number of fermions with wave vector
$\vect k$ and spin $\sigma$ \cite{_spin}, $\epsilon_{\vect k}=\hbar^2\vect k^2/2m$,
$\hbar$ is Plank's constant over $2\pi$ \cite{_hbar},
and $m$ is each fermion's mass. This equation
is valid for \textit{any} nonrelativistic state; it is also valid when we have only
a few fermions.

What happens to the above simple formula
 if a contact interaction between the two components (``spin states") is turned on, 
characterized by a nonzero s-wave scattering length $a$?
The system's wave function now behaves like
\begin{equation}\label{eq:boundary}
\phi=(1/r-1/a)A+O(r)
\end{equation}
if the distance $r$ between two fermions in different spin states is small. Here $A$ depends
on the position of the center of mass of the two fermions, as well as the positions of the remaining fermions.
This system has a superfluid phase with BEC-BCS crossover
\cite{Eagles1969PR,Leggett1980article,_crossover} and other interesting properties, and has
attracted a lot of experimental 
and theoretical work. So such a basic question deserves
an answer.

Because of the interaction, the momentum distribution now in general only decays like $1/k^4$
at large $\vect k$ \cite{_Stringari_momentum}, and the kinetic energy diverges
\cite{_Stringari_momentum}.
In a real system, a contact interaction is impossible, and the divergence is eventually
cut off at $k\sim1/r_0$, where $r_0$ is the range of the interaction
\cite{_r_0}. The interparticle interaction energy also depends sensitively
on the physics at the short scale $r_0$. 
The sum of the two energies (denoted $E_\text{internal}$ below), however,
should be independent of such short-distance
physics, when $r_0$ is much smaller than the other relevant length scales in the problem, including
$\lvert a\rvert$.

We show in this paper that
there is still a simple relation between the energy and the momentum distribution,
which is \textit{independent of the details of the short-range interactions},
except the scattering length $a$ \cite{_Feshbach}:
\begin{equation}\label{eq:energetics}
E_\text{internal}=\frac{\Omega C}{4\pi am}+\lim_{K\rightarrow\infty}\sum_{k<K,\sigma}\frac{k^2}{2m}
\biggl(n_{\vect k\sigma}-\frac{C}{k^4}\biggr),
\end{equation}
where
\begin{equation}\label{eq:C}
C\equiv\lim_{\vect k\rightarrow\infty}
k^4 n_{\vect k\up}=\lim_{\vect k\rightarrow\infty} k^4 n_{\vect k\down},
\end{equation}
$\Omega$ is the volume of space \cite{_volume},
$n_{\vect k\sigma}\equiv \langle c_{\vect k\sigma}^\dagger c_{\vect k\sigma}^{}\rangle$,
and $c_{\vect k\sigma}^{}$ is the standard annihilation operator of a fermion with
wave vector $\vect k$ and spin $\sigma$.
\emph{Equation~\eqref{eq:energetics} holds for any finite-energy states,
no matter few-body or many-body, equilibrium or nonequilibrium,
zero temperature or finite temperature, superfluid state or normal state.
It holds for any populations of the two spin states (balanced or imbalanced).}\cite{_smooth_potential}

In the noninteracting limit, namely $k_Fa\rightarrow 0^{-}$,
$C\propto a^2$ \cite{_noninteracting_limit},
and the first term on the right-hand side of \eqref{eq:energetics} goes to $0$.
Here $k_F$ is the Fermi wave number.
In the unitarity limit, in which $k_Fa\rightarrow\infty$
\cite{_Bertsch}, $C$ approaches some finite value,
and the first term also vanishes.

Some technical remarks about Eq.~\eqref{eq:energetics}:
1) here and throughout this paper, ``infinite momentum'' physically
stands for a momentum scale much lower than $1/r_0$, but much higher than any other
relevant momentum scales in the problem, namely
$1/\lvert a\rvert$, $1/l$, and $1/\lambda_{dB}$, where $l$ is the typical interfermionic
distance, and $\lambda_{dB}$ is the typical de Broglie wave length \cite{_deBroglie};
2) $E_\text{internal}$ is the expectation value of the total energy minus external potential
energy \cite{_internal_energy};
3) the summation over momentum is
understood as $\sum_{\vect k}\equiv \Omega\int[\dif^3k/(2\pi)^3]$,
and there is no infrared divergence. 

In Sec.~\ref{sec:LambdaFunction} a generalized function $\Lambda(\vect k)$,
called the Lambda function, is introduced. Our motivation is to streamline
the formulation of the s-wave contact interaction problem.
Properties of $\Lambda(\vect k)$ are discussed.
It is then used to study the two-body problem.

In Sec.~\ref{sec:selectors} another generalized function, $L(\vect k)$,
is introduced. This and $\Lambda(\vect k)$ span a two dimensional linear
space. A special element in this space, $\eta(\vect k)$, is defined.

In Sec.~\ref{sec:EnergyTheorem} we prove \eqref{eq:energetics} using $\eta(\vect k)$.
Physical implications are discussed, including a possible experimental
test of \eqref{eq:energetics}, and results concerning the pair correlation strength
and the one-body reduced density matrix.

In Sec.~\ref{sec:many_body} we formulate
the whole s-wave contact interaction problem (both few-body and many-body)
in momentum space in a straightforward and unambiguous way,
using the aforementioned generalized functions \cite{_democracy}.
Using this formalism, we rederive \eqref{eq:energetics} in a simpler way,
and solve the three-body problem for the low energy scattering between
a fermion and a weakly bound dimer of fermions.
~

In Sec.~\ref{sec:Summary} we summarize our findings, and discuss the possibility
of extending the ideas contained in our method to some other important physical systems
which appear at first sight to have little to do with the Fermi system considered in this paper.

\section{\label{sec:LambdaFunction}The Lambda Function}
\subsection{Introducing $\Lambda(\vect k)$}
If the interaction has a range $r_0$ that is much shorter
than the other length scales in the problem (including $\lvert a\rvert$), it
can be replaced by the famous pseudopotential \cite{LeeHuangYang1957PR}
\begin{equation}
V=\frac{4\pi a}{m}\delta(\vect r)\frac{\partial}{\partial r}r,
\end{equation}
where $r$ is the distance between two fermions in different spin states,
and the partial derivative is carried out for fixed center-of-mass
location of these two fermions, as well as locations of the remaining particles.
When $r_0\rightarrow0$ but $a$ is kept constant, the pseudopotential becomes exact.
~

To formulate our current problem in momentum space, we perform a Fourier transform. To be concrete,
we first consider two fermions interacting in vacuum. The wave function associated with their
relative motion, $\phi(\vect r)$, satisfies a simple Schr\"{o}dinger equation,
\begin{equation}\label{eq:two_body_1}
E_r\phi=-\frac{\nabla^2}{m}\phi+\frac{4\pi a}{m}\delta(\vect r)
\frac{\partial}{\partial r}(r\phi),
\end{equation}
where $E_r$ is the energy of relative motion. Writing
$\phi(\vect r)=\int\frac{\dif^3k}{(2\pi)^3}\widetilde{\phi}(\vect k)\E^{\I\vect k\cdot\vect r},$
we get
\begin{equation}\label{eq:two_body_2}
\int\frac{\dif^3k}{(2\pi)^3}\biggl(E_r-\frac{k^2}{m}\biggr)\widetilde{\phi}(\vect k)
e^{i\vect k\cdot\vect r}
=\frac{4\pi a}{m}\int\frac{\dif^3k}{(2\pi)^3}\widetilde{\phi}(\vect k)
\delta(\vect r)\frac{\partial}{\partial r}\Bigl(re^{i\vect k\cdot\vect r}\Bigr).
\end{equation}
To proceed, we define
\begin{equation}\label{eq:Lambda_appear}
\delta(\vect r)\frac{\partial}{\partial r}\Bigl(re^{i\vect k\cdot\vect r}\Bigr)
\equiv \delta(\vect r)\Lambda(\vect k),
\end{equation}
where $\Lambda(\vect k)$ is so far an unknown function of $\vect k$.

Clearly
\begin{subequations}
\begin{equation}\label{eq:Lambda_a}
\Lambda(\vect k)=1\text{~~~~(if $k<\infty$)},
\end{equation}
but also, if we multiply both sides of Eq.~\eqref{eq:Lambda_appear}
by $1/k^2$, and integrate over $\vect k$, the left side becomes
$\propto\delta(\vect r)\frac{\partial}{\partial r}(r/r)=0$;
we are thus led to an equality
\begin{equation}\label{eq:Lambda_b}
\int\dif^3k\frac{\Lambda(\vect k)}{k^2}=0,
\end{equation}
in apparent contradiction with Eq.~\eqref{eq:Lambda_a}. This difficulty
has hampered the direct application of the pseudopotential in momentum space.

\textit{Actually there is no contradiction here at all.}

Contradiction can only arise
if we force a \textit{third} equality, namely the integral of $\Lambda(\vect k)/k^2$
over the whole $\vect k$-space is equal to the \textit{limit} of the integral
over a \textit{finite} $\vect k$-space region which expands without bound.
If we insist on both \eqref{eq:Lambda_a} and \eqref{eq:Lambda_b},
as we should, \emph{we must give up this third equality}.

At first sight, this decision is alarming, since the full-space integrals
of \textit{all known functions} are defined in terms of such
a limit. In fact this is the \textit{standard definition} of full-space integrals in mathematical
textbooks. However, if $\Lambda(\vect k)$ is accepted as a special
\textit{generalized function}, $\Lambda(\vect k)/k^2$ does not have to obey this rule.

~

For convenience we postulate one more property for $\Lambda(\vect k)$:
\begin{equation}\label{eq:Lambda_c}
\Lambda(-\vect k)=\Lambda(\vect k),
\end{equation}
whose usefulness will be clear shortly. 
\end{subequations}

Equations~\eqref{eq:Lambda_a}, \eqref{eq:Lambda_b} and \eqref{eq:Lambda_c}
define a generalized function, $\Lambda(\vect k)$, where $\vect k$ is a three-dimensional vector.

Equation~\eqref{eq:Lambda_appear} is then a corollary of such a definition.

\subsection{Mathematical properties of $\Lambda(\vect k)$}
\begin{subequations}
Because $\Lambda(c\vect k)$ (where $c$ is any nonzero finite real constant)
satisfies the same three basic equations as $\Lambda(\vect k)$ itself,
\begin{equation}\label{eq:Lambda_p0}
\Lambda(c\vect k)=\Lambda(\vect k).
\end{equation}
Similarly
\begin{equation}\label{eq:lambda_p0.5}
\Lambda^*(\vect k)=\Lambda(\vect k).
\end{equation}
For any finite constant vector $\vect k_0$,
\begin{equation}\label{eq:Lambda_p1}
\int\dif^3k\frac{\Lambda(\vect k)}{(\vect k-\vect k_0)^2}=0.
\end{equation}
To prove \eqref{eq:Lambda_p1}, we rewrite its left hand side as
\begin{equation*}
\int\dif^3k\Lambda(\vect k)\biggl[\frac{1}{(\vect k-\vect k_0)^2}+\frac{c}{k^2}\biggr]
\end{equation*}
by using \eqref{eq:Lambda_b}. Here $c$ is \emph{any} finite constant. If $c=-1$, the integrand decays
like $1/k^3$ at large $\vect k$, in a given direction. Without Eq.~\eqref{eq:Lambda_c},
we could not determine the integral unambiguously. Using \eqref{eq:Lambda_c}, however,
we can take the \textit{average} between the expression
in the bracket and its spatial inversion, and get
\begin{equation*}
\int\dif^3k\Lambda(\vect k)\biggl[\frac{1}{2(\vect k-\vect k_0)^2}
+\frac{1}{2(-\vect k-\vect k_0)^2}+\frac{c}{k^2}\biggr].
\end{equation*}
If $c=-1$, the integrand now decays like $1/k^4$ at large $\vect k$, and
the integral is dominated by a finite region of the $\vect k$-space (whose size is of the order $k_0$),
so according to Eq.~\eqref{eq:Lambda_a}, the Lambda function can now be dropped.
The resultant \textit{ordinary} integral turns out to be $0$.
Equation~\eqref{eq:Lambda_p1} is thus proved. It follows that
for any finite constant $\vect k_0$ and nonzero finite real constant $c$
\begin{equation}\label{eq:Lambda_p2}
\Lambda(c\vect k-\vect k_0)=\Lambda(\vect k).
\end{equation}
Similarly, it is easy to show, \textit{eg},
\begin{equation}\label{eq:Lambda_p3}
\int\frac{\dif^3k}{(2\pi)^3}~
\frac{\Lambda(\vect k)}{(\vect k-\vect k_0)^2+\alpha^2}=-\frac{\alpha}{4\pi}
~~~~\text{($\alpha\ge0$)},
\end{equation}
which is the $\vect k$-space representation of the simple fact that
the symmetric average of $\E^{-\alpha r+\I\vect k_0\cdot\vect r}/4\pi r$ at $\vect r\rightarrow0$,
excluding the $1/4\pi r$ term, is $-\alpha/4\pi$.
Equation~\eqref{eq:Lambda_p3} is a generalization of \eqref{eq:Lambda_p1}.

A notable corollary of the Lambda function, as illustrated by \eqref{eq:Lambda_p3}, is that
the full-space integral of a function that is \textit{positive} for all
finite $\vect k$'s may still be \textit{negative}.
We will see that this phenomenon is actually \textit{useful} (in Sec.~\ref{subsec:implication}).



The integrals involving the Lambda function still satisfy many familiar rules.
The region of integration, for example,
can be freely divided into some subregions
(with one restriction; see below),
and the total integral equals
the sum of the integrals over these subregions. For example,
\begin{align*}
\int\dif^3k\frac{\Lambda(\vect k)}{k^2}
&=\int_{k<K}\dif^3k\frac{1}{k^2}+\int_{k>K}\dif^3k\frac{\Lambda(\vect k)}{k^2}\\
&=(+4\pi K)+(-4\pi K)=0,
\end{align*}
for any finite $K\ge 0$.
In the ball region, $\vect k$ is finite and $\Lambda(\vect k)=1$,
while in the external region the integral is identical with that
of $\Lambda(\vect k)\theta(k-K)/k^2$ over the whole $\vect k$-space,
which can be computed with the same method as is used to derive Eq.~\eqref{eq:Lambda_p1}.
Here for simplicity we separate the subregions
with a sphere, but other arbitrary shapes of the boundaries
are equally permissible, \textit{provided} that
infinite momentum is contained by \textit{only one} subregion.
\end{subequations}

Like the delta function, $\Lambda(\vect k)$ can freely participate in operations like addition,
multiplication, integration, and Fourier transformation. It also commutes
with other functions and quantum mechanical state vectors and operators, ie,
$\Lambda(\vect k)X=X\Lambda(\vect k)$, where $X$ may or may not depend on $\vect k$.

\subsection{The Fourier transform of $\Lambda(\vect k)$}

The Fourier transform of $\Lambda(\vect k)$ is
\begin{equation}\label{eq:lambda}
\lambda(\vect r)\equiv\int\frac{\dif^3k}{(2\pi)^3}\Lambda(\vect k)\E^{\I\vect k\cdot\vect r}.
\end{equation}

\begin{subequations}
Since $\Lambda(\vect k)$ is both even and real, so is $\lambda(\vect r)$:
\begin{equation}\label{eq:lambda_p1}
\lambda(-\vect r)=\lambda(\vect r),
\end{equation}
\begin{equation}\label{eq:lambda_p2}
\lambda^*(\vect r)=\lambda(\vect r).
\end{equation}

It follows from Eqs.~\eqref{eq:Lambda_p0} and \eqref{eq:lambda} that
\begin{equation}\label{eq:lambda_p2.5}
\lambda(c\vect r)=\lvert c\rvert^{-3}\lambda(\vect r),
\end{equation}
a property reminiscent of the three-dimensional delta function.
Here $c$ is any nonzero finite real constant.

Equation~\eqref{eq:Lambda_p1} can be expressed in $\vect r$-space as
$\int\dif^3r\lambda(\vect r)\E^{\I\vect k_0\cdot\vect r}/r=0$. Expanding
this result in powers of $k_0$, we get
\begin{equation}\label{eq:lambda_p3}
\int\dif^3r\lambda(\vect r)\frac{1}{r}=0,
\end{equation}
\begin{equation}\label{eq:lambda_p4}
\int\dif^3r\lambda(\vect r)\hat{\vect r}=0.
\end{equation}

The fact that $\Lambda(\vect k)$ is equal to
1 for any finite $\vect k$ leads to
\begin{equation}\label{eq:lambda_p5}
\int\dif^3r\lambda(\vect r)f(\vect r)=f(0),
\end{equation}
for any function $f$ which is \textit{finite} and \textit{continuous} in a region containing
$\vect r=0$. This entails, in particular, that
\begin{equation}\label{eq:lambda_p6}
\lambda(\vect r)=0~~~~\text{(if $\vect r\neq0$).}
\end{equation}

From Eqs.~\eqref{eq:lambda_p3}, \eqref{eq:lambda_p4}, and \eqref{eq:lambda_p5}, we get
\begin{equation}\label{eq:lambda_p7}
\int\dif^3r\lambda(\vect r)\frac{g(\vect r)}{r}=0,
\end{equation}
for any function $g$ which is \textit{smooth} in a region containing $\vect r=0$.
This equation becomes obvious if we write $g(\vect r)=A+\vect B\cdot\vect r+O(r^2)$.

According to Eq.~\eqref{eq:lambda_p6}, all the integrals involving $\lambda(\vect r)$,
like the ones showed above, can be restricted to a neighborhood of the origin without affecting
their values. This is an important similarity between
$\lambda(\vect r)$ and $\delta(\vect r)$. The two generalized functions mainly differ in two
aspects: 1) $\int\dif^3r\lambda(\vect r)/r=0$
but $\int\dif^3r\delta(\vect r)/r$ is divergent;
2) $\int\dif^3r\lambda(\vect r)\hat{\vect r}=0$
but $\int\dif^3r\delta(\vect r)\hat{\vect r}$ is undefined.

$\lambda(\vect r)$ is related to the operator $\delta(\vect r)(\partial/\partial r)r$:
\begin{equation}\label{eq:lambda_p8}
\int\dif^3r\delta(\vect r)\frac{\partial}{\partial r}[rf(\vect r)]
=\int\dif^3r\lambda(\vect r)f(\vect r),
\end{equation}
for any function $f$ for which $\int\dif^3r\delta(\vect r)\frac{\partial}{\partial r}[rf(\vect r)]$
is well-defined.

Finally, if we integrate both sides of Eq.~\eqref{eq:Lambda_appear} over $\vect r$,
and use Eq.~\eqref{eq:lambda_p8}, we get
\begin{equation}\label{eq:lambda_p9}
\int\dif^3r\lambda(\vect r)\E^{\I\vect k\cdot\vect r}=\Lambda(\vect k),
\end{equation}
which is just the Fourier transformation from the $\vect r$-space to the $\vect k$-space.
This equation is consistent with \eqref{eq:lambda}, because the Lambda function is even.
\end{subequations}

It will be clear that $\Lambda(\vect k)$ and $\lambda(\vect r)$ allow for
a much more flexible treatment of the s-wave
contact interaction problem
than the operator $\delta(\vect r)(\partial/\partial r)r$.

\subsection{Two-body problem}
The right-hand side of \eqref{eq:two_body_2} equals
$\int\frac{\dif^3k'}{(2\pi)^3}\Lambda(\vect k')\widetilde{\phi}(\vect k')\delta(\vect r)$
according to \eqref{eq:Lambda_appear}.
Expanding $\delta(\vect r)$ in terms of plane waves, and comparing both
sides of Eq.~\eqref{eq:two_body_2}, we obtain the $\vect k$-space representation of \eqref{eq:two_body_1},
\begin{equation}\label{eq:two_body_3}
E_r\widetilde{\phi}(\vect k)=\frac{k^2}{m}\widetilde{\phi}(\vect k)
+\frac{4\pi a}{m}\int\frac{\dif^3k'}{(2\pi)^3}
\Lambda(\vect k')\widetilde{\phi}(\vect k')
\end{equation}
or, equivalently,
\begin{equation*}\label{eq:two_body_4}
E_r\widetilde{\phi}(\vect k)=\frac{k^2}{m}\widetilde{\phi}(\vect k)
+\frac{4\pi a}{m\Omega}\sum_{\vect k'}
\Lambda(\vect k')\widetilde{\phi}(\vect k').
\end{equation*}

To familiarize ourselves with this formalism, we consider a simple exact
solution to \eqref{eq:two_body_3}, namely the bound state ($E_r<0$).
Let $f$ denote the second term on the right-hand side of \eqref{eq:two_body_3}.
Solving \eqref{eq:two_body_3} formally, we get
\begin{equation}\label{eq:phik}
\widetilde{\phi}(\vect k)=-f/(k^2/m-E_r)~~~(f\ne0)
\end{equation}
which is smooth for all $\vect k$. Substituting \eqref{eq:phik} back into
the definition of $f$, and using \eqref{eq:Lambda_p3},
we get $f=fa\sqrt{-mE_r}$. So $a>0$, $E_r=-1/ma^2$, and
$\widetilde{\phi}(\vect k)\propto1/(k^2+a^{-2})$, in perfect agreement with the established wisdom
\cite{Leggett2001RMP}.

We can easily extend the above approach to three or more particles
(see Sec.~\ref{subsec:3body} for an illustration).

\section{\label{sec:selectors}Short-Range Selectors}

\subsection{$L(\vect k)$}

If the connection between $\int\dif^3k$ and $\lim_{K \rightarrow\infty}
\int_{\lvert\vect k\rvert<K}\dif^3k$ is not universal, nothing can prevent the existence
of another generalized function, $L(\vect k)$, defined as follows:
\begin{subequations}
\begin{equation}\label{eq:L_a}
L(\vect k)=0\text{~~~~(if $k<\infty$)},
\end{equation}
\begin{equation}\label{eq:L_b}
\int\frac{\dif^3k}{(2\pi)^3}\frac{L(\vect k)}{k^2}=1,
\end{equation}
\begin{equation}\label{eq:L_c}
L(-\vect k)=L(\vect k).
\end{equation}
\end{subequations}
As generalized functions, $\Lambda(\vect k)$ and $L(\vect k)$ can both be approached by ordinary
functions. If $\vect k$ is a wave vector, $L(\vect k)$'s dimension is length, while $\Lambda(\vect k)$
is dimensionless.







\subsection{Properties of the $L$ function}


\begin{subequations}
\begin{equation}\label{eq:L_p0}
\int\frac{\dif^3k}{(2\pi)^3}f(\vect k)L(\vect k)=\lim_{\vect k\rightarrow\infty}k^2f(\vect k)
\end{equation}
for any ordinary function $f(\vect k)$.

We now list some other properties of $L(\vect k)$;
their proofs are similar to those in Sec.~\ref{sec:LambdaFunction}.
\begin{equation}\label{eq:L_p1}
L(c\vect k)=\lvert c\rvert^{-1}L(\vect k)
\end{equation}
for any real constant $c\ne0$.
\begin{equation}\label{eq:L_p2}
L^*(\vect k)=L(\vect k).
\end{equation}
\begin{equation}
L(\vect k-\vect k_0)=L(\vect k)
\end{equation}
for any constant vector $\vect k_0$.
\end{subequations}
An integral involving $L(\vect k)$, like that involving $\Lambda(\vect k)$,
can be freely divided into many subintegrals
(\textit{provided} that infinite momentum is contained by \textit{only one}
subregion). $L(\vect k)$ can also freely participate in various operations, and it commutes
with other objects. 
~

Now turn to the coordinate representation of the $L$ function,
\begin{equation}\label{eq:l}
l(\vect r)\equiv\int\frac{\dif^3k}{(2\pi)^3}L(\vect k)\E^{\I\vect k\cdot\vect r}.
\end{equation}
\begin{subequations}

For any ordinary function $f(\vect r)$,
\begin{equation}\label{eq:l_p1}
\int\dif^3r l(\vect r)f(\vect r)=\lim_{\vect r\rightarrow0}
4\pi r f(\vect r).
\end{equation}
Other properties of $l(\vect r)$ are listed below.
\begin{equation}
l(\vect r)=0~~\text{(if $\vect r\ne0$)}.
\end{equation}
\begin{equation}\label{eq:l_p2}
l^*(\vect r)=l(-\vect r)=l(\vect r).
\end{equation}
\begin{equation}\label{eq:l_p4}
l(c\vect r)=\lvert c\rvert^{-2}l(\vect r)
\end{equation}
for any real constant $c\ne0$. Equation~\eqref{eq:l_p4} is consistent
with the dimension of $l(\vect r)$, namely [length]$^{-2}$.
\begin{equation}\label{eq:l_p5}
\int\dif^3rl(\vect r)\E^{-\I\vect k\cdot\vect r}=L(\vect k).
\end{equation}

\end{subequations}

\subsection{\label{subsec:selectors}Introducing the short-range selectors}


Let $f_1(\vect k)$ be any ordinary function satisfying $\int\frac{\dif^3k}{(2\pi)^3}f_1(\vect k)=1$.
Let $f_2(\vect k)=1/k^2+r'(\vect k)$ be any ordinary function satisfying
$\int\frac{\dif^3k}{(2\pi)^3}r'(\vect k)=0$.
Let $s_1(\vect k)=\Lambda(\vect k)$ and $s_2(\vect k)=L(\vect k)$. We have
\begin{equation}\label{eq:dual}
\int\frac{\dif^3k}{(2\pi)^3}s_i^*(\vect k)f_j(\vect k)=\delta_{ij}.
\end{equation}

$f_1(\vect k)$ and $f_2(\vect k)$ span a two-dimensional linear space $\mathcal{F}$.
$\Lambda(\vect k)$ and $L(\vect k)$ span another two-dimensional linear space $\mathcal{S}$.
Equation \eqref{eq:dual} states that $\mathcal{F}$ and $\mathcal{S}$ are \emph{dual linear spaces}.

We shall call the elements of $\mathcal{S}$ (short-range) selectors,
because for any function $f(\vect k)=\sum_ic_if_i(\vect k)$, we can selectively extract the coefficient
$c_i$ using an element in $\mathcal{S}$: $\int\frac{\dif^3k}{(2\pi)^3}s_i^*(\vect k)f(\vect k)=c_i$.

All these functions can be Fourier-transformed to the $\vect r$-space.
Let $\widetilde{s}_i(\vect r)=\int\frac{\dif^3k}{(2\pi)^3}s_i(\vect k)
\E^{\I\vect k\cdot\vect r}$, $\widetilde{f}_i(\vect r)=\int\frac{\dif^3k}{(2\pi)^3}
f_i(\vect k)\E^{\I\vect k\cdot\vect r}$.
\begin{equation}
\int\dif^3r\widetilde{s}_i^*(\vect r)\widetilde{f}_j(\vect r)=\delta_{ij},
\end{equation}
which represents a group of nontrivial equations:
\begin{equation}\begin{array}{cc}
\int\dif^3r\lambda(\vect r)=1, &
\int\dif^3r\lambda(\vect r)/(4\pi r)=0,\\
\int\dif^3r l(\vect r)=0, &
\int\dif^3r l(\vect r)/(4\pi r)=1.
\end{array}
\end{equation}
Also,
\begin{equation}
\widetilde{s}_i(\vect r)=0~~\text{(if $\vect r\ne0$)}.
\end{equation}

Each short-range selector corresponds to a linear functional which extracts
a short-range property of an ordinary function $\widetilde{f}(\vect r)$.

At this point, we have completed a tool which will free us from ill-defined ultraviolet divergences,
for the two-component Fermi gas with s-wave contact interaction \cite{_spin}.
In this formalism, no \textit{ad hoc} large momentum cut-offs or dimensional regularizations are needed.

Here we have presented our approach in a generic form, so that it \textit{might}
be possible to extend it to other physical systems involving contact
interactions. For the other systems we may need more than two linearly independent short-range
selectors. For instance, we may need \emph{three} independent selectors when the two components
of the Fermi gas with s-wave contact interaction have a mass ratio exceeding 13.6, so that
there is Efimov effect which introduces an additional parameter for the interaction
\cite{Petrov2003PRA,Braaten2004}.

\subsection{The $\eta$-selector}
The $\eta$-selector is a particular element in $\mathcal{S}$:
\begin{subequations}
\begin{align}
\eta(\vect k)&\equiv\Lambda(\vect k)+\frac{L(\vect k)}{4\pi a},\label{eq:eta_k}\\
\widetilde{\eta}(\vect r)&\equiv\int\frac{\dif^3k}{(2\pi)^3}
\eta(\vect k)\E^{\I\vect k\cdot\vect r}=\lambda(\vect r)+\frac{l(\vect r)}{4\pi a}.
\end{align}\end{subequations}
This selector will play a crucial role in the s-wave contact interaction problem,
because it selectively annihilates the relative wave function of two particles with such interaction:
\begin{subequations}
\begin{equation}\label{eq:eta_p1}
\int\dif^3r\widetilde{\eta}(\vect r)[1/r-1/a+O(r)]=0.
\end{equation}
On the other hand,
\begin{equation}\label{eq:eta_p2}
\int\dif^3r\widetilde{\eta}(\vect r)f(\vect r)=f(0),
\end{equation}
for any ordinary function $f(\vect r)$ that is continuous in the neighborhood of the origin,
so $\widetilde{\eta}(\vect r)$ behaves like the
delta function for nonsingular functions. More properties of the $\eta$-selector are listed below.
\begin{equation}\label{eq:eta_p3}
\int\dif^3r\widetilde{\eta}(\vect r)\hat{\vect r}=0.
\end{equation}
\begin{equation}\label{eq:eta_p4}
\widetilde{\eta}(\vect r)=0~~~~\text{(if $\vect r\neq0$)}.
\end{equation}
\begin{equation}\label{eq:eta_p5}
\widetilde{\eta}^*(\vect r)=\widetilde{\eta}(-\vect r)=\widetilde{\eta}(\vect r).
\end{equation}
\begin{equation}\label{eq:eta_p6}
\eta(\vect k)=1~~~~\text{(if $k<\infty$)}.
\end{equation}
\begin{equation}\label{eq:eta_p7}
\int\frac{\dif^3k}{(2\pi)^3}\frac{\eta(\vect k)}{k^2}=\frac{1}{4\pi a}.
\end{equation}
\begin{equation}\label{eq:eta_p8}
\int\frac{\dif^3k}{(2\pi)^3}\eta(\vect k)f(\vect k)=
\frac{c}{4\pi a}+\lim_{K\rightarrow\infty}\int_{\lvert\vect k\rvert<K}
\frac{\dif^3k}{(2\pi)^3}\big[f(\vect k)-\frac{c}{k^2}\big],
\end{equation}
where $f(\vect k)$ is an ordinary function and $c=\lim_{\vect k\rightarrow\infty}k^2f(\vect k)$.
\end{subequations}

\section{\label{sec:EnergyTheorem}Energy Theorem}

\subsection{Mathematical formulation}
\textbf{Theorem.}
If the system of fermions of equal mass $m$ populating two spin states with s-wave contact interaction
and scattering length $a$ is in a smooth external potential $V_\text{ext}(\vect r)$, and is in a state
$\hat{\rho}=\sum_{i=1}^\infty \alpha_i\ket{\phi_i}\brac{\phi_i}$
(where $\bracket{\phi_i}{\phi_j}=\delta_{ij}$, $\alpha_i\ge0$, and $\sum_i\alpha_i=1$) satisfying two conditions:
firstly each $\ket{\phi_i}$ is a linear combination of energy eigenstates with coefficients of the combination
decaying sufficiently fast at large energy such that the wave function $\phi_i$ in the coordinate representation
has no singularities other than those introduced by the interfermionic interaction, and secondly
the probability $\alpha_i$ decays sufficiently fast at large $i$ such that
\begin{equation}\label{eq:Ci}
C=\sum_i\alpha_iC_i,
\end{equation}
where $C$ [defined by \eqref{eq:C}] is associated with the state $\hat{\rho}$, and $C_i$ associated with
$\ket{\phi_i}$ \cite{_Ci}, then the system's energy expectation value is
\begin{equation}\label{eq:EnergyTheorem}
E=\sum_{\vect k\sigma}\eta(\vect k)\frac{k^2}{2m}n_{\vect k\sigma}
+\sum_{\sigma}\int\dif^3rV_\text{ext}(\vect r)\rho_{\sigma}(\vect r).
\end{equation}
Here $\eta(\vect k)$ is defined in the previous section, $n_{\vect k\sigma}$ is the momentum distribution,
and $\rho_\sigma(\vect r)$ is the spatial density distribution.

\textit{Proof.}
The second term on the right-hand side of \eqref{eq:EnergyTheorem} is trivial. We will concentrate
on the first term, $E_\text{internal}$, which is physically the sum of the total kinetic energy
and the interfermionic interaction energy, both of which are divergent in the zero-range
interaction limit. However, $E_\text{internal}$ can be unambiguously determined
in this limit (with $a\ne0$ fixed), using the pseudopotential method.

Let us first consider the case in which the system is in a pure state $\ket{\phi}$ having
exactly $N$ fermions in the spin up state and $M$ fermions in the spin down state:
\begin{multline}
\ket{\phi}=\frac{1}{N!M!}\int\dif^3r_1\cdots\dif^3r_N\dif^3s_1\cdots\dif^3s_M
\phi(\vect r_1, \cdots, \vect r_N, \vect s_1, \cdots, \vect s_M)\\\times
\psi_{\up}^\dagger(\vect r_1)\cdots\psi_{\up}^\dagger(\vect r_N)
\psi_{\down}^\dagger(\vect s_1)\cdots\psi_{\down}^\dagger(\vect s_M)\ket{0},
\end{multline}
where $\ket{0}$ is the particle vacuum, and $\psi_\sigma(\vect r)$ is the standard
fermion annihilation operator at spin state $\sigma$ and spatial location $\vect r$.
The wave function $\phi(\vect r_1, \cdots, \vect r_N, \vect s_1, \cdots, \vect s_M)$
is completely antisymmetric under the exchange of any two fermions
in the same spin state; it has also been properly normalized:
\begin{equation}
\frac{1}{N!M!}\int\dif^3r_1\cdots\dif^3r_N\dif^3s_1\cdots\dif^3s_M
\lvert\phi(\vect r_1, \dots, \vect r_N, \vect s_1, \dots, \vect s_M)\rvert^2=1,
\end{equation}
so that $\bracket{\phi}{\phi}=1$.

The \textit{only role} of the pseudopotential is to \textit{exactly cancel}
the delta function singularities arising when the kinetic energy operator
(which is just a $3N+3M$-dimensional Laplace operator divided by $-2m$)
acts on the wave function. These delta functions arise when two fermions
with opposite spins come together. To understand this, we may examine
Eq.~\eqref{eq:two_body_1} closely; note also that this cancellation mechanism
is carried over to the arbitrary-body cases, with or without external potentials.
The internal energy of the fermions is therefore given by a simple expression,
\begin{equation}\label{eq:E_int}
-2mE_\text{internal}=\frac{1}{N!M!}\lim_{\epsilon\rightarrow0}\int_{\mathcal{D}(\epsilon)}
\dif^{3N+3M}R~\phi^*(\vect R)\nabla^2\phi(\vect R),
\end{equation}
where $\vect R$ is the shorthand for the $3N+3M$ coordinates of the fermions,
and $\nabla^2$ is the $3N+3M$-dimensional Laplace operator.
$\mathcal{D}(\epsilon)$ is a subset of the $3N+3M$-dimensional configuration space,
\textit{excluding} the regions in which any two fermions with opposite spins
have a distance \textit{less than} $\epsilon$.

Now define another quantity $X$,
\begin{multline}\label{eq:X}
-2mX\equiv\frac{1}{N!M!}\int\dif^{3N+3M}R\dif^3t
\,\widetilde{\eta}(\vect t)\phi^*(\vect r_1\cdots\vect r_N\vect s_1\cdots\vect s_M)\\
\times\nabla_{\vect t}^2\Bigl[\sum_{i=1}^{N}\phi(\vect r_1\cdots\vect r_{i-1},
\vect r_i+\vect t, \vect r_{i+1}\cdots\vect r_N\vect s_1\cdots\vect s_M)\\
+\sum_{j=1}^{M}\phi(\vect r_1\cdots\vect r_N\vect s_1\cdots\vect s_{j-1},
\vect s_j+\vect t, \vect s_{j+1}\cdots\vect s_M)\Bigr],
\end{multline}
and we want to prove that $X=E_\text{internal}$. To do so, we divide the
$3N+3M$-dimensional $\vect R$-space in this new integral into $\mathcal{D}(\epsilon)$
and $\mathcal{I}(\epsilon)$, where $\mathcal{D}(\epsilon)$ is the same as above,
and $\mathcal{I}(\epsilon)$ is complementary to $\mathcal{D}(\epsilon)$.
Clearly, in the subregion $\mathcal{D}(\epsilon)$,
the integral over $\vect R$ is finite and continuous,
and the result is a continuous function of $\vect t$ for $t<\epsilon$;
then, according to the basic property of $\widetilde{\eta}(\vect t)$,
it can be treated as the delta function, and we immediately see the integral
is exactly equal to the integral in $E_\text{internal}$.

We then only need to show that the integral in $\mathcal{I}(\epsilon)$
approaches zero as $\epsilon \rightarrow0$. Since the volume of $\mathcal{I}(\epsilon)$
is proportional to $\epsilon^3$ when $\epsilon$ is sufficiently small,
$\mathcal{I}(\epsilon)$ can be further divided into $NM$ subregions,
in one of which $\lvert\vect r_1-\vect s_1\rvert<\epsilon$. All the
other $NM-1$ subregions are the same as this, due to fermionic symmetry.
We have omitted subregions with volumes of higher orders in $\epsilon$.
In the thermodynamic limit, however, even when $\epsilon$ is very small,
we can still have many pairs of fermions, and the distance between two fermions
in each of these pairs is smaller than $\epsilon$. But we note that these pairs
are far apart if $\epsilon$ is small, so they can be treated \textit{independently}.

So now we only consider the case in which $\lvert\vect r_1-\vect s_1\rvert<\epsilon$,
which is representative of the general situation. In this case, only two terms
in the big bracket on the right side of Eq.~\eqref{eq:X} have the possibility of
making contributions to the total integral which do not approach zero. One of them
is the term in which $\vect r_1$ is replaced by $\vect r_1+\vect t$, the other being the one
in which $\vect s_1$ is replaced by $\vect s_1+\vect t$. We only discuss the first term,
since the logic is the same for the second one.

To treat this term we make a coordinate transformation: $\vect r=\vect r_1-\vect s_1$
and $\vect r_0=(\vect r_1+\vect s_1)/2$, and represent $\vect r_2, \cdots,
\vect r_N, \vect s_2, \cdots, \vect s_M$ with a single $3N+3M-6$ dimensional vector
$\vect R'$. We then first do integral over $\vect r$ (and $r<\epsilon$),
then do the integral over $\vect t$, and finally integrate over $\vect r_0$ and $\vect R'$.

Expanding $\phi(\vect R',\vect r_0,\vect r)$ in this case as $A(\vect R',\vect r_0)
(1/r-1/a)+O(r)$
(according to the short range boundary condition), we write our target integral in
the form $$Y=\int\dif^{3N+3M-6}R'\dif^3r_0\int\dif^3t\widetilde{\eta}(\vect t)
\nabla_{\vect t}^2K(\vect R', \vect r_0, \vect t),$$ where we have omitted the constant
coefficient since it is irrelevant to our question, and
\begin{multline*}
K(\vect R', \vect r_0, \vect t)\equiv
\int_{r<\epsilon}\dif^3r\big[A^*(\vect R', \vect r_0)(1/r-1/a)+O(r)\big]\\\times
\big[A(\vect R', \vect r_0+\vect t/2)(1/\lvert\vect r+\vect t\rvert-1/a)
+O(\lvert\vect r+\vect t\rvert)\big].
\end{multline*}

$K$ should be expanded in powers of the small
$\vect t$, before we can carry out the integral
over $\vect t$; $t$ is regarded as much smaller than $\epsilon$,
because $\widetilde{\eta}(\vect t)$ is zero for any nonzero $\vect t$.
In such an expansion, any term which contains a factor
$\epsilon$ raised to any positive power should be omitted. Also,
any term which is of the order $t^3$ or higher should be omitted, since
it contributes nothing to the integral
$\int\dif^3t\widetilde{\eta}(\vect t)\nabla_{\vect t}^2[\cdot]$.
We then have $K\sim c_1(t/2-t^2/6a)+\vect c_2\cdot\hat{\vect t}t^2$,
where $c_1=-4\pi \lvert A(\vect R', \vect r_0)\rvert^2$,
and $\vect c_2=-\pi A^*(\vect R', \vect r_0)\nabla_{\vect r_0}A(\vect R', \vect r_0)$.
But now $\nabla_{\vect t}^2K\sim c_1(1/t-1/a)+4\vect c_2\cdot\hat{\vect t}$,
and according to the short range properties of $\widetilde{\eta}(\vect t)$,
the integral over $\vect t$ vanishes.

The above analysis shows that the contribution to $X$ from $\mathcal{I}(\epsilon)$
approaches zero as $\epsilon\rightarrow 0$, so $X=E_\text{internal}$.

Equation \eqref{eq:X} can be easily rewritten in the second quantized
form: $$-2mX=\brac{\phi}\sum_\sigma\int\dif^3r\dif^3t\widetilde{\eta}(\vect t)
\nabla_{\vect t}^2\psi_\sigma^\dagger(\vect r)\psi_\sigma(\vect r+\vect t)\ket{\phi}.$$
Expanding $\psi_\sigma(\vect r)=\Omega^{-1/2}\sum_{\vect k_1}c_{\vect k_1\sigma}^{}
\exp(i\vect k_1\cdot\vect r)$
and similarly for $\psi_\sigma(\vect r+\vect t)$, and carrying out the integration over
$\vect r$ and $\vect t$, we get $$-2mX=\brac{\phi}\sum_{\vect k\sigma}\eta(\vect k)
(-k^2)c_{\vect k\sigma}^\dagger c_{\vect k\sigma}^{}\ket{\phi}.$$ So
$E_\text{internal}=\sum_{\vect k\sigma}\eta(\vect k)
(k^2/2m)\brac{\phi}c_{\vect k\sigma}^\dagger c_{\vect k\sigma}^{}\ket{\phi}$.

So far, we have proved \eqref{eq:EnergyTheorem} for a pure quantum state, with fixed numbers
of fermions in the two spin states.

If the pure state is not an eigenstate of particle numbers in the two spin states,
we can expand it as a superposition of such eigenstates; since the interaction
conserves the number of fermions in each spin state, the
off-diagonal matrix elements of the Hamiltonian between these eigenstates are zero, and 
the theorem remains valid.

If the system is in a mixed state, described by the density operator $\hat{\rho}$, and
the different eigenstates of $\hat{\rho}$ independently satisfy the theorem,
then the statistical ensemble of these
states still satisfies the theorem, provided that \eqref{eq:Ci} holds.
~

Using Eq.~\eqref{eq:eta_p8} to reexpress the first term on the right-hand side
of \eqref{eq:EnergyTheorem}, we get \eqref{eq:energetics}.

From the many-body wave function, we can also prove Eq.~\eqref{eq:C} by expanding
$\langle\psi_\sigma^\dagger(\vect r)\psi_\sigma^{}(\vect r+\vect t)\rangle$ at small $\vect t$.
In this expansion there is a singular term (proportional to $t$ but independent of $\hat{\vect t}$)
that is independent of $\sigma$:
\begin{gather}
\langle\psi_\sigma^\dagger(\vect r)\psi_\sigma^{}(\vect r+\vect t)\rangle=\rho_\sigma(\vect r)
-\frac{C(\vect r)t}{8\pi}+\frac{\I}{\hbar}\vect p_\sigma(\vect r)\cdot\vect t+O(t^2),\\
C(\vect r)\equiv\frac{16\pi^2}{(N-1)!(M-1)!}\int\dif^{3N+3M-6}R'\lvert A(\vect R',\vect r)\rvert^2,
\end{gather}
where $A(\vect R', \vect r_0)$ is defined in the above proof. Consequently $\vect n_{\vect k\sigma}$
decays like $C/k^4$ at large $\vect k$, and
\begin{equation}\label{eq:Cr_C}
\Omega C=\int C(\vect r)\dif^3r\equiv\mathcal{I}.
\end{equation}
We can also show a result for the \emph{pair correlations}:
\begin{equation}\label{eq:PairCorrelation}
\big\langle\hat{\rho}_\up(\vect r-\vect t/2)\hat{\rho}_\down(\vect r+\vect t/2)\big\rangle
=\frac{C(\vect r)}{16\pi^2}\Big(\frac{1}{t^2}-\frac{2}{at}\Big)+O(t^0)
\end{equation}
at small $\vect t$,
where $\hat{\rho}_\sigma(\vect r)\equiv\psi_\sigma^\dagger(\vect r)\psi_\sigma^{}(\vect r)$.


\subsection{Other proofs}

For those who are not satisfied with the proof of the energy theorem
with the $\eta$-selector, we have in principle at least two other
proofs. One of them is almost identical with the proof presented; instead
of directly using the $\eta$-selector to annihilate some terms, we may
do a detailed analysis of the behavior of 
$\langle\psi_\sigma^\dagger(\vect r)\psi_\sigma^{}(\vect r+\vect t)\rangle$
at small $\vect t$, and, besides the term proportional to $t$ (independent of
the direction of $\vect t$), we also need to analyze
the term proportional to $t^2$ (also independent of $\hat{\vect t}$).
There is also a term proportional to the inner product of
$t^2\hat{\vect t}$ and a direction vector, but it contributes nothing
if we do the symmetric integral in momentum space (as is done
in Eq.~\eqref{eq:energetics}).

The other proof is inspired by what Eq.~\eqref{eq:energetics} tells us.
Instead of studying an idealized s-wave contact interaction model,
we may study a short-range ($r_0$) attractive interaction potential, and
fine-tune the depth of the potential to achieve a specified scattering length
($\lvert a\rvert\gg r_0$). Then, we can divide the total internal
energy in two pieces. One of them is the integral of kinetic energy up to
a momentum scale $K$, where $K$ is much smaller than $1/r_0$ but
much larger than the other characteristic momentum scales in the problem.
The other piece is the kinetic energy integral from $K$ to
much higher than $1/r_0$, \textit{plus} all the interfermionic interaction
energy. It is then possible to show that the second piece can be
expressed in terms of $C\approx K^4n_{\vect K\sigma}^{}$:
approximately $[\pi/(2a)-K]\Omega C/(2\pi^2m)+O(1/K)$, with other errors
that vanish in the limit $r_0\rightarrow0$ (but the scattering length
is kept constant). In the two-body case, this approach can work out
without much difficulty. In the many-body cases, it is however
tricky to give a rigorous proof; but
there is a \textit{heuristic} physical picture: at large $K$,
we are effectively probing those fermions each of which is close
to another fermion (with distance $\sim 1/K$), and for each of such
fermion pair, two-body physics is a good approximation.

\subsection{Physical Implications\label{subsec:implication}}

One can measure the momentum distribution $n_{\vect k\sigma}$ experimentally.
A well-known method is to suddenly switch off
\textit{both} the interaction between fermions \textit{and}
the external confinement potential, to allow the fermionic cloud to expand ballistically.
Because particles with different momenta move at different velocities, eventually
the spatial density distribution will reflect the momentum distribution.
(The spatial distribution can be measured with some imaging
technique.) See, \textit{eg}, Ref.~\cite{Greiner2005PRL}.

\begin{figure}\begin{center}
\includegraphics{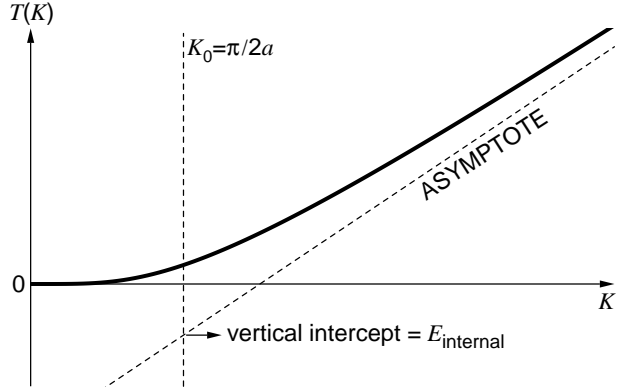}
\caption{\label{fig:energy_theorem}Partial kinetic energy $T(K)$ versus $K$.
The vertical intercept of the asymptote at $K_0=\pi/2a$ equals the
internal energy, according to \eqref{eq:TestIt}.}\end{center}
\end{figure}

From the measured $n_{\vect k\sigma}$, one can compute a partial kinetic energy,
by summing contributions from all $\vect k$'s up to scale $K$:
\begin{equation}\label{eq:Ek}
T(K)\equiv\sum_{\lvert\vect k\rvert<K,\sigma}
\frac{\hbar^2k^2}{2m}n_{\vect k\sigma},
\end{equation}
and plot $T(K)$ versus $K$.
Because of the nonzero $C$ \cite{_nonzeroC}
(in the $C/k^4$ tail of the momentum distribution),
$T(K)$ does not approach any finite value for $K\ll1/r_0$,
but instead, it approaches a straight line with a positive slope,
which is just the \textit{asymptote} of the curve $T(K)$.
What Eq.~\eqref{eq:energetics} states is simply that
\textit{the internal energy is exactly equal to the
vertical intercept of such an asymptote at $K_0\equiv\pi/(2a)$}.
See Fig.~\ref{fig:energy_theorem} for illustration.
More specifically,
\begin{equation}\label{eq:TestIt}
T(K)=E_\text{internal}+\frac{\hbar^2\Omega C}{2\pi^2m}
\left(K-\frac{\pi}{2a}\right) +O(1/K),
\end{equation}
for $K$ much larger than the other momentum scales in the problem
but much smaller than $1/r_0$. This result is \textit{universally
valid}, as is stressed in Sec.~\ref{sec:introduction}.

The $O(1/K)$ term on the right-hand side of \eqref{eq:TestIt}
is related to the second order singularity of the relative wave function
of two fermions with opposite spins. The first order singularity is
like $1/r$ and the second order is like $r$. The interference between
these two orders gives rise to a term $\propto 1/K$ at large $K$.

If we plot $T(K)$ all the way up to the scale $1/r_0$,
the above asymptote behavior breaks down, and the curve eventually
approaches a horizontal line, associated with the total kinetic energy,
which is much larger than the internal energy.

It is simple mathematics to confirm the energy theorem in the case of
an isolated bound state of two fermions, whose momentum distribution is \cite{Leggett2001RMP}
$$n_{\vect k\sigma}=\frac{C}{(\lvert \vect k-\vect k_0\rvert^2+a^{-2})^2},~~~C=\frac{8\pi}{a\Omega},$$
where $2\hbar\vect k_0$ is the total momentum. In this case
the first term on the right-hand side of
Eq.~\eqref{eq:energetics} equals $+2\hbar^2/(ma^2)$, and the second term
equals $-3\hbar^2/(ma^2)+(\hbar k_0)^2/m$. The total
energy is $-\hbar^2/(ma^2)+(\hbar k_0)^2/m$, simply a sum of the binding energy \cite{Leggett2001RMP}
and the center-of-mass kinetic energy. Note that the total energy is negative if $k_0<a^{-1}$.

Here we see that the symbol $\eta(\vect k)$ has to break
the conventional law of integral (that the integral over the whole $\vect k$-space equals the limit
of integrals over finite regions of this space), otherwise
it would be impossible for the first term on the right-hand side of
Eq.~\eqref{eq:EnergyTheorem} to generate any \textit{negative number},
since \textit{$\eta(\vect k)$ is equal to $+1$ for any finite $\vect k$
and the momentum distribution is also everywhere positive}. 
\textit{The negative binding energy is associated with
this purely positive momentum distribution.} We could of
course reject this modified notion of integration, and stick to
Eq.~\eqref{eq:energetics} only, but then
the beautiful structure of the problem, as shown by \eqref{eq:EnergyTheorem}, would
be obscured. 
~

The energy theorem takes a particularly simple form in the unitarity
limit: when $a\rightarrow\infty$, the internal energy is just the intercept
of the aforementioned asymptote at \textit{zero momentum}.

The quantum few-body problem in the unitarity limit in a harmonic trap is
preliminarily studied in \cite{Tan0412764}; we may also study the momentum
distributions of these few-body systems and relate them to the energies,
using Eq.~\eqref{eq:TestIt}.

Equations \eqref{eq:Cr_C} and \eqref{eq:PairCorrelation} imply that
the expectation of the number of pairs of fermions with
diameters smaller than a \textit{small} distance $s$ is
\begin{equation}\label{eq:Npair}
N_\text{pair}=\frac{\Omega Cs}{4\pi}
\end{equation}
if $s$ is still much larger than $r_0$.
More specifically, the expectation of the number of pairs of fermions
located in a volume element $\dif^3r$, with diameters smaller than a small distance $s$ is
\begin{equation}
\dif N_\text{pair}=\frac{C(\vect r)s}{4\pi}\dif^3r.
\end{equation}
For these reasons, we shall call $C(\vect r)$ \emph{local contact intensity},
$\mathcal{I}=\int C(\vect r)\dif^3r=\Omega C$ \emph{integrated} contact intensity,
and $C$ \emph{average} contact intensity (over volume $\Omega$).

In a trapped Fermi gas, when the Thomas-Fermi approximation
is valid, we may also have an approximate concept
of local momentum distribution, valid within the limit set by Heisenberg
uncertainty principle. Here we are using the term
``Thomas-Fermi approximation'' in its broadest sense, namely
the fermionic cloud is divided into many portions: each portion is \textit{both} so small
that it is roughly uniform, \textit{and}
so large that it can be treated as in the thermodynamic limit,
so that all the thermodynamic quantities are meaningful.

Even when the Thomas-Fermi approximation is invalid (eg in few-body systems),
the local contact intensity $C(\vect r)$ is still an \textit{exact} concept. The resolution $\delta k$ of the
momentum distribution within a local spatial region is of the order the inverse size of the region.
In the large momentum part, where we can use a large unit for momentum,
$\delta k$ is negligible, and a tail $C(\vect r)/k^4$ is well defined.

We also have a concept of \textit{local internal energy density} $\epsilon_\text{internal}$,
inspired by an intermediate step in the proof of the energy theorem:
\begin{equation}\label{eq:local_internal}
\epsilon_\text{internal}(\vect r)\equiv-\sum_\sigma\int\dif^3s~
\widetilde{\eta}(\vect s)\frac{\hbar^2\nabla_{\vect s}^2}{2m}
\langle\psi_\sigma^\dagger(\vect r)
\psi_\sigma^{}(\vect r+\vect s)\rangle,
\end{equation}
and the total internal energy is exactly
\begin{equation}
E_\text{internal}=\int\epsilon_\text{internal}(\vect r)\dif^3r.
\end{equation}

The above results concerning local pair correlations and local energy densities
can help us to address a fundamental problem for a finite number of fermions,
namely how to construct a systematically
improvable perturbation theory, with the Thomas-Fermi
approximation as the zeroth order approximation. 
The problem is important for two reasons: 1) ultracold
atomic Fermi gases realized so far typically
contain hundreds of thousands of atoms, for which the Thomas-Fermi approximation
has detectable errors, and
2) if we want to use a quantum few-body system to simulate the many-body
thermodynamic limit (see, eg, \cite{Tan0412764}), finite size
corrections must be taken into account. In fact few-body calculations are \textit{all that we can do}.
BCS theory, for example, is all about the two-fermion correlations
in the presence of a many-body \textit{mean field}. 
Quantum Monte-Carlo simulations are restricted to a small number of particles.
In fact we can not solve Schr\"{o}dinger equation for more than a handful
of particles (except for the noninteracting systems or some very special systems).
~

In the thermodynamic limit, the behavior of $C$ or the complete momentum distribution
on the entire $(-1/k_Fa, T/T_F)$ plane (for balanced populations of the two spin states)
is worth studying. Here $T$ is temperature, and
$T_F=\hbar^2 k_F^2/2m k_B$ the Fermi temperature.
Since the energy does not change smoothly across the phase transition line, the energy theorem
implies that the momentum distribution also has some unsmooth change.

Another implication of the energy theorem concerns a common dynamic process:
the Fermi gas is initially confined in a trap, and then the trap potential
is suddenly turned off (but the scattering length is not changed), so
that the gas expands in the presence of interactions.
During the expansion, the contact intensity continuously decreases,
in such a way that the vertical intercept [at $K_0=\pi/(2a)$] of the
\textit{asymptote} of the function $T(K)$ remains constant because of
energy conservation. This is
a \textit{constraint} on the intantaneous momentum distributions.

If the Fermi gas is in a motion (or if we are looking at a
\textit{local portion} of the expanding cloud in the local approximation),
and the momentum distribution is roughly symmetric around
$\vect k=\vect k_0$, it is better
to use the following alternative formula to evaluate the internal energy
\begin{equation}
E_\text{internal}=\frac{\hbar^2\Omega C}{4\pi am}+\lim_{K\rightarrow\infty}
\sum_{\lvert\vect k-\vect k_0\rvert<K,\sigma}\left[\frac{\hbar^2k^2}{2m}n_{\vect k\sigma}-
\frac{\hbar^2C}{2m(\vect k-\vect k_0)^2}\right],
\end{equation}
and $C=\lim_{\vect k-\vect k_0\rightarrow\infty}
\lvert\vect k-\vect k_0\rvert^4n_{\vect k\sigma}$.
These two formulas are equivalent to Eqs.~\eqref{eq:energetics} and \eqref{eq:C}, in the
contact-interaction limit. But now $k_0$ can exceed $1/r_0$ without leading to
any problems, provided that the typical \textit{relative} energy of these fermions is
much smaller than $\hbar^2/(mr_0^2)$.

\section{\label{sec:many_body}Many-Body Problem in Momentum Space}

\subsection{Basic Formulation}

Many-body theories are often studied in momentum space, because
in the thermodynamic limit, we often have translational symmetry,
and the exploitation of this symmetry in momentum space simplifies
many things.

The s-wave contact interaction problem, however, lacks a satisfactory
momentum representation to date. The peculiar contact interaction causes
ultraviolet divergence problems, which some authors deal with by using
concepts like ``bare'' coupling constants and renormalized ones; the
shortcoming of these approaches
is that the bare constants are ill-defined divergent quantities, and the
sum of them and the divergent counterterms are ambiguous.

Here we write down the momentum formulation of the problem, using the
short-range selectors.

We start from the coordinate space. The many-body Hamiltonian is
\begin{equation}
H=H_\text{internal}+\int\dif^3r\sum_\sigma V_\text{ext}(\vect r)
\psi_\sigma^\dagger(\vect r)\psi_\sigma^{}(\vect r),
\end{equation}
\begin{align}
H_\text{internal}&=-\frac{1}{2m}\int\dif^3r\sum_\sigma\psi_\sigma^\dagger(\vect r)\nabla^2
\psi_\sigma^{}(\vect r)\notag\\
&\quad\,+\frac{4\pi a}{m}
\int\dif^3r\dif^3r'\lambda(\vect r')\psi_\up^\dagger(\vect r)
\psi_\down^\dagger(\vect r)\psi_\down^{}(\vect r-\vect r'/2)
\psi_\up^{}(\vect r+\vect r'/2).\label{eq:H_r}
\end{align}
Note the similarity between the above equation and Eqs.~(1) and (2)
of Ref.~\cite{LeeHuangYang1957PR}, where the hard-sphere \textit{Bose}
gas is studied; but in our context, in the contact
interaction limit, the pseudopotential Hamiltonian is exact.
Any \textit{right-hand side eigenstate} $\ket{\phi}$ of the Hamiltonian
(satisfying $H\ket{\phi}=E\ket{\phi}$),
or any linear combinations of such eigenstates (with an upper bound in energy),
automatically satisfy the short range boundary condition
\begin{equation}\label{eq:phi_r}
\int\dif^3r'\widetilde{\eta}(\vect r')\psi_\down
(\vect r-\vect r'/2)\psi_\up(\vect r+\vect r'/2)\ket{\phi}
=0.
\end{equation}
All the state vectors which satisfy this boundary condition
form a subspace of the Hilbert space, and we call it the
\textit{physical subspace}, and represent it with $\mathcal{P}$
\cite{_vacuum}.


Equations \eqref{eq:H_r} and \eqref{eq:phi_r} can be rewritten in the momentum space:
\begin{equation}\label{eq:H_k}
H_\text{internal}=\sum_{\vect k\sigma}\frac{k^2}{2m}c_{\vect k\sigma}^\dagger
c_{\vect k\sigma}^{}
+\frac{4\pi a}{m\Omega}\sum_{\vect q\vect k\vect k'}\Lambda(\vect k')
c_{\vect q/2+\vect k\up}^\dagger c_{\vect q/2-\vect k\down}^\dagger
c_{\vect q/2-\vect k'\down}^{}c_{\vect q/2+\vect k'\up}^{},
\end{equation}
\begin{equation}\label{eq:phi_k}
\sum_{\vect k}\eta(\vect k)c_{\vect q/2-\vect k\down}c_{\vect q/2+\vect k\up}
\ket{\phi}=0~~~~\text{(for any $\vect q$)}.
\end{equation}
All the ultraviolet divergence problems
disappear, when $\Lambda(\vect k')$ is restored
in the Hamiltonian. No divergent bare constants,
no \textit{ad hoc} regularizations, no renormalizations. Everything
can be formulated \textit{simply} and \textit{exactly}.

We now describe how to compute the expectation value of $H_\text{internal}$ under any state
$\ket{\phi}\in\mathcal{P}$. The rule is very simple:
\begin{multline}\label{eq:H_expectation}
\langle H_\text{internal}\rangle=\sum_{\vect k}
\Bigl\{\sum_\sigma\frac{k^2}{2m}\langle c_{\vect k\sigma}^\dagger
c_{\vect k\sigma}^{}\rangle\\
+\frac{4\pi a}{m\Omega}\sum_{\vect q\vect k'}\Lambda(\vect k')
\langle c_{\vect q/2+\vect k\up}^\dagger c_{\vect q/2-\vect k\down}^\dagger
c_{\vect q/2-\vect k'\down}^{}c_{\vect q/2+\vect k'\up}^{}\rangle\Bigr\},
\end{multline}
that is, the two terms should be grouped in the above way. Now the energy is finite,
so the summation over $\vect k$ is convergent and the summand decays faster than $1/k^3$
at large $\vect k$. Consequently, we can insert the $\eta$-selector,
according to \eqref{eq:eta_p8}:
\begin{multline*}
\langle H_\text{internal}\rangle=\sum_{\vect k}\eta(\vect k)
\Bigl\{\sum_\sigma\frac{k^2}{2m}\langle c_{\vect k\sigma}^\dagger
c_{\vect k\sigma}^{}\rangle\\
+\frac{4\pi a}{m\Omega}\sum_{\vect q\vect k'}\Lambda(\vect k')
\langle c_{\vect q/2+\vect k\up}^\dagger c_{\vect q/2-\vect k\down}^\dagger
c_{\vect q/2-\vect k'\down}^{}c_{\vect q/2+\vect k'\up}^{}\rangle\Bigr\}.
\end{multline*}
Distribute $\eta(\vect k)$ to the two terms, we get
\begin{multline*}
\langle H_\text{internal}\rangle=\sum_{\vect k\sigma}\eta(\vect k)
\frac{k^2}{2m}\langle c_{\vect k\sigma}^\dagger c_{\vect k\sigma}^{}\rangle\\
+\frac{4\pi a}{m\Omega}\sum_{\vect q\vect k\vect k'}\eta(\vect k)\Lambda(\vect k')
\brac{\phi} c_{\vect q/2+\vect k\up}^\dagger c_{\vect q/2-\vect k\down}^\dagger
c_{\vect q/2-\vect k'\down}^{}c_{\vect q/2+\vect k'\up}^{}\ket{\phi}/\bracket{\phi}{\phi},
\end{multline*}
but the second term must vanish, since the Hermitian conjugate of
Eq.~\eqref{eq:phi_k} is $\sum_{\vect k}\eta(\vect k)\brac{\phi}
c_{\vect q/2+\vect k\up}^\dagger c_{\vect q/2-\vect k\down}^\dagger=0$.
Therefore
\begin{equation*}
\langle H_\text{internal}\rangle=\sum_{\vect k\sigma}\eta(\vect k)
\frac{k^2}{2m}\langle c_{\vect k\sigma}^\dagger c_{\vect k\sigma}^{}\rangle,
\end{equation*}
which is precisely the energy theorem.

It is tempting to use this approach as the simplest way to derive the energy
theorem. However it is not explained why the terms should be grouped in the
way as in Eq.~\eqref{eq:H_expectation}, even though it is a very natural grouping.
\textit{It is the logical proof of the energy theorem in the last section
that supports this way of grouping the terms}.

The inner product of Eq.~\eqref{eq:phi_k} with any quantum state is zero.
\textit{This gives us many useful identities}. The simplest of them is
\begin{equation}
\sum_\vect k\eta(\vect k)\langle
c_{-\vect k\down}c_{\vect k\up}\rangle=0,
\end{equation}
which is a constraint on the pairing amplitude in the superfluid state.
We have also seen the constraint on the
two-body reduced density matrix, and we exploited it
to rederive the energy theorem above. \textit{Similar constraints
are present on the three-body, four-body, ..., reduced density matrices.}

The internal Hamiltonian \eqref{eq:H_k}, although not Hermitian in the whole
Hilbert space, \textit{is} Hermitian in the physical subspace. This can be
easily shown from the energy theorem, which indicates that the expectation
value of $H_\text{internal}$ under any state in $\mathcal{P}$ is real.
For any $\ket{\phi_1}, \ket{\phi_2}\in\mathcal{P}$,
and any angle $\theta$, $\ket{\phi_1}+\E^{\I\theta}\ket{\phi_2}$ is also
in $\mathcal{P}$, and the expectation values of $H_\text{internal}$ under all these states are
real. It then follows that $\brac{\phi_1}H_\text{internal}\ket{\phi_2}
=\brac{\phi_2}H_\text{internal}\ket{\phi_1}^*$.

One might worry about the divergence of $a$ in the unitarity
limit, in which the second term on the right-hand side of \eqref{eq:H_k} appears
ill-defined. We can show, however, that there is no real problem here.
Before $a$ becomes divergent, we should use the physical subspace condition
to replace $\Lambda(\vect k')$ with $\Lambda(\vect k')+j\eta(\vect k')$, where $j$ is an
arbitrary constant. Choosing $j=-1$ we get
\begin{equation}
H_\text{internal}\ket{\phi}=\sum_{\vect k\sigma}\frac{k^2}{2m}c_{\vect k\sigma}^\dagger
c_{\vect k\sigma}^{}\ket{\phi}
-\frac{1}{m\Omega}\sum_{\vect q\vect k\vect k'}L(\vect k')
c_{\vect q/2+\vect k\up}^\dagger c_{\vect q/2-\vect k\down}^\dagger
c_{\vect q/2-\vect k'\down}^{}c_{\vect q/2+\vect k'\up}^{}\ket{\phi},
\end{equation}
and the second term is now explicitly well-behaved in the unitarity limit.
This equation is valid for \textit{any} scattering length, no matter
finite or divergent. To calculate the inner product of this
equation with any state vector, we should use Eq.~\eqref{eq:L_p0}
to convert the integral over $\vect k'$ to a limit at large $\vect k'$.

Note that in the unitarity limit the $\eta$-selector is equal to the
$\Lambda$-selector, making many expressions particularly simple.

\subsection{Few-body physics: an example\label{subsec:3body}}

\textit{All the few-body physics} is clearly contained by our second-quantized
formulation.

To show that the formalism presented in this paper is a streamlined
\textit{working method}
(and not only a formal framework), and to show that few-body physics
and many-body physics can be treated with this very \textit{same} formalism,
we demonstrate a simple example, the low energy scattering of a fermionic dimer
and a free fermion, in which case $a>0$ necessarily. This calculation can be
compared with Ref.~\cite{Petrov2005PRA} in which the same problem is studied.

It is desirable to work in momentum space because of spatial translational
symmetry. The system is in a quantum state
\begin{equation}\label{eq:phi_3body}
\ket{\phi}=\sum_{\vect k_1\vect k_2\vect k_3}
\phi_{\vect k_1\vect k_2\vect k_3}
c_{\vect k_1\up}^\dagger c_{\vect k_2\up}^\dagger c_{\vect k_3\down}^\dagger\ket{0},
\end{equation}
where the wave function is antisymmetric under the exchange of $\vect k_1$ and
$\vect k_2$, and is zero if $\vect k_1+\vect k_2+\vect k_3\neq0$. Substituting
\eqref{eq:phi_3body} into Schr\"{o}dinger equation $H\ket{\phi}=E\ket{\phi}$,
where $H$ is given by \eqref{eq:H_k} (assuming $V_\text{ext}=0$), we get,
\begin{multline}\label{eq:3bodySchrodinger}
\left(\frac{k_1^2+k_2^2+k_3^2}{2m}-E\right)\phi_{\vect k_1\vect k_2\vect k_3}\\
+\frac{4\pi a}{m\Omega} \biggl[
\sum_{\vect k_1'\vect k_3'}\Lambda(\vect k_1'-\vect k_3')
\phi_{\vect k_1'\vect k_2\vect k_3'}
+\sum_{\vect k_2'\vect k_3'}\Lambda(\vect k_2'-\vect k_3')
\phi_{\vect k_1\vect k_2'\vect k_3'}\biggr]=0.
\end{multline}
We define
\begin{equation}
f_{\vect k_2}\equiv\frac{1}{\Omega}\sum_{\vect k_1'\vect k_3'}
\Lambda(\vect k_1'-\vect k_3')\phi_{\vect k_1'\vect k_2\vect k_3'},
\end{equation}
which is a function of just the \textit{length} of $\vect k_2$,
in the case of low energy s-wave scattering.
If the relative momentum of the dimer and the fermion approaches $0$,
$E=-1/(ma^2)<0$, and Eq.~\eqref{eq:3bodySchrodinger}
can be formally solved:
\begin{equation}\label{eq:3body2}
\phi_{\vect k_1\vect k_2\vect k_3}=\frac{4\pi a(f_{\vect k_1}-f_{\vect k_2})}
{(k_1^2+k_2^2+k_3^2)/2+1/a^2},
\end{equation}
for $\vect k_1+\vect k_2+\vect k_3=0$.
Substituting this back to the definition of $f_{\vect k_2}$, and writing
$\vect k_2=\vect k$,
$\vect k_1'=-\vect k/2+\vect k'$, and $\vect k_3'=-\vect k/2-\vect k'$,
we get
\begin{equation}
f_{\vect k}^{}=4\pi a\int\frac{\dif^3k'}{(2\pi)^3}\Lambda(\vect k')
\frac{f_{-\vect k/2+\vect k'}^{}-f_{\vect k}^{}}{k'^2+3k^2/4+a^{-2}},
\end{equation}
where the integral of the second term on the right-hand side can be immediately
evaluated, using \eqref{eq:Lambda_p3}. In the first term,
however, $\Lambda(\vect k')$ can be dropped,
because the integrand decays faster than
$1/k'^3$ for $\vect k'\rightarrow\infty$. If this were not the case,
$f_\vect k$ would have to decay like $1/k$ or even slower at large $k$,
and according to Eq.~\eqref{eq:3body2}, $\phi_{\vect k_1\vect k_2\vect k_3}$
would have to decay like $1/K^3$ or slower, when $k_1$, $k_2$ and $k_3$ are all
of the order $K\rightarrow\infty$,
making the wave function unnormalizable at short distances \cite{_normalization}.
We thus get
\begin{equation}\label{eq:f_k}
\bigl(\sqrt{3k^2/4+a^{-2}}-a^{-1}\bigr)f_{\vect k}
+4\pi\int\frac{\dif^3k'}{(2\pi)^3}
\frac{f_{\vect k'}}{k'^2+\vect k\cdot\vect k'+k^2+a^{-2}}=0,
\end{equation}
where $\vect k'$ has been shifted.

The Fourier transform of $f_\vect k$ is proportional
to the relative wave function of the fermion and the dimer at large distances,
and must be of the form $1-a_{fd}/r$ for $r\rightarrow\infty$.
Here $a_{fd}$ is the fermion-dimer scattering length, whose value we shall determine.
Consequently, $f_{\vect k}$ is of the form $(2\pi)^3\delta(\vect k)
-4\pi a_{fd}/k^2$ plus regular terms at small $\vect k$.

Let $f_{\vect k}\equiv F(t)$, where $t\equiv (ka)^2$ is dimensionless.
Equation~\eqref{eq:f_k} is easily simplified to
\begin{equation}
\biggl(\sqrt{1+\frac{3t}{4}}-1\biggr)F(t)
+\frac{1}{\pi\sqrt{t}}
\int_{0}^{\infty}F(t')~\mathrm{arctanh}\frac{\sqrt{tt'}}{1+t+t'}\dif t'=0.
\end{equation}
It can be shown that $F(t)\propto -\pi\delta(t)/\sqrt{t}+(a_{fd}/a)/t$ plus
a \textit{smooth} function for all $t\ge0$ (including the neighborhood of $t=0$),
and that $F(t)\propto c_0 t^{\mu_0/2}+ c_1 t^{(\mu_0-1)/2}$ plus
higher order terms at $t\rightarrow\infty$, where
$$\mu_0=-4.16622197664779257337, \text{~and~~} c_1/c_0=0.30268913080233667524.$$
Exploiting these properties, we can discretize the integral equation in an appropriate
way, and solve it extremely accurately. The resultant fermion-dimer scattering length
turns out to be
\begin{equation}
a_{fd}=1.1790662349~a.
\end{equation}
To make an extremely accurate
prediction for a \textit{real} system (eg, ultracold atoms),
we have to take various corrections to the idealized s-wave contact interaction
into account, a topic beyond the scope of this paper.

\section{\label{sec:Summary}Summary and Outlook}

To solve a paradox arising in the formulation of the s-wave contact interaction
problem of ultracold fermions in momentum space, we modified the conventional notion
of integrals. This unexpectedly led us to a simple relation between the energy and
the momentum distribution of these fermions.

The new notion is that while the integral of an ordinary function over the whole space
is always equal to the limit of integrals over finite regions of space, some generalized
functions do not obey this rule. For them, certain full-space integrals are specified beforehand,
in a way that is compatible with their other properties and with conventional mathematics.

Our approach is \textit{very different} from the ones used by some people in
some relativistic quantum field theories (in which they force
integrals like $\int\dif^4k k^{-2}=0$), even though one might form this false impression
if one only reads Sec.~\ref{sec:LambdaFunction} of this paper.
In our approach, for example, $\Lambda(\vect k)/k^2$ and $\eta(\vect k)/k^2$
are equal for all finite $\vect k$'s, but their full-space integrals are different.
The infinitely many possibilities form a \textit{linear space}, called
the \textit{selector space}, in which each point corresponds to certain values
of some integrals. Also, in our approach,
integrals of ordinary functions which are divergent remain divergent forever
(for example, $\int\dif^4k k^{-2}=\infty$),
and the generalized functions and ordinary ones \textit{coexist}, forming a coherent system.

We discussed the physical implications of our calculations. In particular, if we
measure the momentum distribution of fermions with s-wave contact interactions,
and plot the result in a certain way, a simple asymptote appears, and
the height of such an asymptote
at a certain horizontal coordinate is equal to the internal energy.
This result is the generalization of a simple property of noninteracting
particles, in the context of strongly interacting ultracold quantum gases with resonant
interactions ($\lvert a\rvert\gg r_0$).

We also discussed the application of the results to confined quantum gases,
and found that the energy of the system can be extracted from
the equal-time \textit{one-body reduced density matrix} only. This is of course completely
different from the density functional theory of electrons,
in which the explicit form of the functional is \textit{unknown}.

We formulated the s-wave contact interaction problem of ultracold fermions in a
simple and coherent way, and can now freely transform the problem between coordinate
space and momentum space. This is a streamlined formalism for studying many concrete problems,
including few-body ones.

Full applications of this formalism
to the many-body problem are a topic of future research. Because of the
peculiar structure of this system, we can not predict whether there will be further surprises
awaiting us. The short-range boundary condition of the many-body wave function,
essentially the definition of the scattering length, appears to be a highly nontrivial constraint.
Can we construct a series of approximations to reduce discrepancies with this
constraint in a progressive way, and to minimize the system's free energy
at the same time? It is possible.

The BCS wave function is of course not consistent with this constraint.
For example, $\sum_{\vect k'}\eta(\vect k') \brac{\phi_\text{BCS}} c_{\vect q/2+\vect k\up}^\dagger
c_{\vect q/2-\vect k\down}^\dagger c_{\vect q/2-\vect k'\down}
c_{\vect q/2+\vect k'\up}\ket{\phi_\text{BCS}}\not\equiv0$ at $\vect q\neq 0$.

I hope to extend some essential ideas in this paper to \textit{all} systems
with contact interactions, including the Standard Model of particle
physics. If there are $R$ independent ``renormalized'' constants that must be
determined by experiments (or by a higher-energy theory) in a low-energy effective theory,
is the dimension of the associated selector space equal to $R+1$? We have this tentative guess,
because in the nonrelativistic s-wave contact interaction problem considered in this
paper, we have only one ``renormalized'' constant, the scattering length,
and our selector space is two-dimensional. Another conjecture is that
the selector spaces of those other theories may have some nontrivial algebraic
properties; for example, the momentum translation of one selector may lead to
mixtures with other ones. It may also be interesting to investigate how these selectors
evolve (\textit{inside the selector space}) under the continuous scaling transformation,
to see whether or how the conventional notion of renormalization group flow
is incorporated. In the two-dimensional selector space described in this paper,
$\Lambda(\vect k)$ is unchanged if $\vect k$ is rescaled,
but $L(\vect k)$ undergoes a simple scaling transformation, and a generic selector
[linear combination of $\Lambda(\vect k)$ and $L(\vect k)$] generally changes
after a scaling transformation.

If such a universal approach is realized, certain developments will follow.
The most important of them will probably be related to some nonperturbative properties
of interacting quantum fields. Even if we can not determine all the properties
quantitatively, we may still find some exact relations and/or qualitative features.

\ack
The author thanks E.~Braaten, C.~Chin, L.~I.~Glazman, M.~Greiner, T.~L.~Ho, K.~Levin,
and D.~S.~Petrov for comments. This work was supported, in part, by DOE Grant
No. DE-FG02-00ER41132.

\bibliographystyle{elsart-num}
\bibliography{energy}

\begin{thebibliography}{10}
\bibitem{_hbar}
We let $\hbar=1$ for conciseness, and the term ``wave vector" is used
  interchangeably with ``momentum". $\hbar$ is restored when contact with
  experiments is made.
\bibitem{_crossover}
The BEC-BCS crossover is not restricted to a Fermi gas with resonant contact
  interaction characterized by the s-wave scattering length, although the
  latter system is the most popular, and this paper is restricted to this
  minimal model.
\bibitem{_Feshbach}
Since we are assuming that the interaction is characterized by the scattering
  length only, the theory in this paper is \textit{not} directly applicable to
  Fermi gases near \textit{narrow} Feshbach resonances, where there can be a
  length scale $R^*$ which may be comparable to or larger than $\lvert a\rvert$
  \cite{Petrov2004PRL_Rstar}. But to broad Feshbach resonances, where $R^*$ is
  small, our calculations are directly applicable.
\bibitem{_volume}
Periodic boundary condition is used. In the case of a uniform gas, $\Omega$ is
  the volume of the uniform system; in the case of a confined gas, $\Omega$ is
  any volume which greatly exceeds the actual spatial extension of the gas.
\bibitem{_smooth_potential}
The system is assumed to be in a \emph{smooth} external potential (may be
  zero), so that the $C/k^4$ tail of $n_{\vect k\sigma}$ is solely caused by
  interparticle interactions.
\bibitem{_noninteracting_limit}
To appreciate this general feature, it is instructive to consider two
  interacting fermions confined in a box.
\bibitem{_deBroglie}
A convenient definition of the ``typical" de Broglie wave length is
  $1/m\lambda_{dB}^2\sim\epsilon_1$, where $\epsilon_1$ is the total internal
  energy (kinetic energy plus interfermionic interaction energy) divided by the
  number of fermions, in the center-of-mass frame of the Fermi gas.
\bibitem{_Ci}
The mixed state $\hat{\rho}=\sum_\vect kC_\up(k^2+a^{-2})^{-2}\ket{\vect
  k}\brac{\vect k}$, where $\ket{\vect k}=c_{\vect k\up}^\dagger\ket{0}$ is a
  single-fermion momentum eigenstate, violates both the second equality in
  \eqref{eq:C} and \eqref{eq:Ci}. This counterexample shows the necessity of
  the condition that the probability $\alpha_i$ (in
  $\hat{\rho}=\sum_i\alpha_i\ket{\phi_i}\brac{\phi_i}$) decay sufficiently fast
  at large $i$.
\bibitem{_nonzeroC}
In the contact-interaction model, there are many ``interactionless" states,
  where any two fermions with opposite spins have precisely zero probability to
  be in the s-wave orbital channel, so that $C=0$ even though $a\ne0$. For
  many-body systems, such states are difficult to realize experimentally,
  unless some unexpected technique is invented.
\bibitem{_vacuum}
The vacuum state and the single-fermion states also belong to the physical
  subspace. Any linear combination of any two states which belong to
  $\mathcal{P}$ also belongs to $\mathcal{P}$ and is physically allowed.
\end{thebibliography}

\end{document}